\documentclass[11pt]{article}

\usepackage[margin=1in]{geometry}
\usepackage{newpxtext}
\usepackage{amsmath,amsthm}
\usepackage{newpxmath}
\usepackage{mathtools}
\usepackage{microtype}
\usepackage[colorlinks=true,linkcolor=blue,citecolor=blue,urlcolor=blue]{hyperref}
\usepackage{cleveref}
\usepackage{enumitem}
\usepackage{booktabs,tabularx,array}

\usepackage{tikz}
\usetikzlibrary{patterns}

\usepackage[skins]{tcolorbox}
\tcbuselibrary{breakable}

\newtheorem{theorem}{Theorem}
\newtheorem{lemma}[theorem]{Lemma}
\newtheorem{corollary}{Corollary}

\newcommand{\Oh}{O^*}

\title{Faster Quantum Algorithms for TSP and More}
\author{
Masayuki Miyamoto
\\University of Tsukuba, Japan
 }
\date{}

\begin{document}
\maketitle

\begin{abstract}
The traveling salesman problem (TSP) is a canonical NP-hard optimization problem. For general edge weights, the classical dynamic programs of Bellman and of Held and Karp solve TSP on \(n\) cities in \(O^*(2^n)\) time. To date, no $\Oh(c^n)$-time classical algorithm is known for any $c<2$. Combining Grover search with the dynamic programming technique, Ambainis et al. (SODA 2019) introduced a quantum algorithm with running time $O^*(1.7274^n)$, which is still the best known quantum time bound. 

In this work, we improve the running time to $O^*(1.70961^n)$.
The core idea of our algorithm is to combine the framework of Ambainis et al. with the extremal set-system framework, which was recently used for classical space-time tradeoffs of TSP by Jabal Ameli, Nederlof and Wang (FOCS 2026) and Dallant and Kozma (FOCS 2026). To obtain our upper bound, we decompose the optimal tour into four consecutive subpaths of different sizes, for which, based on the idea of set-systems, the optimal values are computed using dynamic programming with small success probability and use amplitude amplification. Using the same idea, we also give a new quantum space-time tradeoff for TSP, improving the previous results of Caroppo et al. (ESA 2026) and Dallant (arXiv:2607.12374). Finally, we obtain the same quantum upper bounds for several other related problems.
\end{abstract}

\section{Introduction}\label{sec:introduction}

Given a weighted graph, the traveling salesman problem (TSP) asks for
a minimum-weight cycle visiting every vertex exactly once. Whereas brute-force search requires $\Oh(n!)$ running time, the algorithms of Bellman~\cite{bellman} and Held and Karp~\cite{hk} from 1962 solve TSP in time $O(n^22^n)=\Oh(2^n)$. Throughout, \(\Oh\) hides factors polynomial
in the input length.

\paragraph{Classical algorithms for special cases of TSP.}
When every edge has unit cost, every Hamiltonian cycle has cost \(n\),
so solving TSP is equivalent to deciding whether a Hamiltonian cycle exists.
For undirected graphs, Bj{\"o}rklund~\cite{bjorklund-ham} gave a
randomized \(O^\ast(1.657^n)\)-time algorithm for this decision problem.
For directed graphs, Li~\cite{li-ham} showed that the exact number of
Hamiltonian cycles can be computed in \(2^{n-\Omega(\sqrt n)}\) time.
Koana and Kumabe~\cite{kk-ham} subsequently gave a one-sided randomized
\(O^\ast(1.9133^n)\)-time algorithm for deciding whether a directed
Hamiltonian cycle exists, the first to achieve a fixed exponential base
below \(2\) for this decision problem on arbitrary directed graphs.

For weighted TSP, several algorithms exploit restrictions on the edge
costs. On directed graphs with nonnegative integer costs bounded by a
polynomial in \(n\), Bj{\"o}rklund~\cite{bjorklund-below} obtained a
running time of \(2^{n-\Omega(\sqrt{n/\log n})}\). On undirected graphs
with positive integer edge costs, Bj{\"o}rklund~\cite{bjorklund-ham}
gave a randomized algorithm that finds the minimum tour cost in
\(O^\ast(W\,1.657^n)\) time, where \(W\) is the sum of the edge costs.
Stoian~\cite{stoian-doubling} obtained randomized \(O^\ast(1.66^n)\)
time for undirected TSP when the set \(A\) of nonnegative integer edge
costs satisfies \(\lvert\{a+b:a,b\in A\}\rvert=O(\lvert A\rvert)\).

Restrictions on the graph can also give a time base below \(2\), even
with arbitrary edge costs. For asymmetric TSP, Bj{\"o}rklund, Husfeldt,
Kaski, and Koivisto~\cite{bhkk-degree} obtained such a bound when the
underlying undirected graph has bounded maximum degree. For undirected
TSP with arbitrary real edge costs, Fang, Guo, Lu, and Yao~\cite{fgly}
gave a deterministic algorithm with time base below \(2\) whenever
the graph contains an independent set of size at least \(\alpha n\),
for any fixed \(3/7<\alpha\le 1/2\). On bipartite graphs their
algorithm runs in \(O^\ast((1+2\sqrt{2})^{n/2})\) time, giving a
time base below \(1.957\). An earlier randomized algorithm of
Nederlof~\cite{nederlof-bip} solved bipartite TSP in
\(O(1.9999^n)\) time assuming matrix multiplication exponent \(2\) (a November 2026 preprint by OpenAI~\cite{openai-matrix-exponent}
reports that the exponent is \(9/4\)).

\paragraph{Quantum algorithms and space--time tradeoffs.}
In contrast to special cases, for general TSP, no classical algorithm is known to run in \(O^\ast(c^n)\)
time for any constant \(c<2\). In 2019, Ambainis et al.~\cite{abi}
combined classical precomputation on small subsets with nested quantum
minimum finding. They obtained \(O^\ast(1.7274^n)\) time for (directed and undirected) TSP on
\(n\) cities.

Like Bellman--Held--Karp, the quantum TSP algorithm of Ambainis et al.\
stores path costs for exponentially many sets of cities. How fast can
TSP be solved with less memory? Let \(S\) and \(T\) denote the
exponential bases of space and time, so that the respective bounds are
\(S^n 2^{o(n)}\) and \(T^n 2^{o(n)}\). Bellman--Held--Karp gives
\((S,T)=(2,2)\), while the divide-and-conquer approach of Gurevich and
Shelah gives polynomial space and \((S,T)=(1,4)\)~\cite{bellman,hk,gs,dk}.
Combining the two gives intermediate tradeoffs with \(ST=4\).
A first improvement of $ST$ was given by Koivisto and Parviainen~\cite{kp}, achieving \(ST\le 3.9271\).

A different idea has recently improved this tradeoff beyond 3.9271. Jabal Ameli, Nederlof, and Wang~\cite{anw} and, independently, Dallant and Kozma~\cite{dk} run the Bellman–Held–Karp recurrence only over a set system, that is, a family of city sets. The restricted recurrence finds every tour whose first $t$ cities form a member of the set system for every $t$. The sets of first $t$ cities for $t=1,2,\ldots,n$ form a maximal chain, that is, a nested sequence of city sets of sizes $1,2,\ldots,n$, and every maximal chain arises from exactly one visiting order in this way. A useful set system therefore has few members but contains many maximal chains. The set systems of~\cite{anw,dk} disprove a conjecture of Johnson, Leader, and Russell~\cite{jlr} on which set systems of a given size contain the most maximal chains. With this approach, the resulting algorithms use less space and more time than Bellman–Held–Karp, and the product $ST$ drops to below 3.572~\cite{dk} and then to 3.1861~\cite{anw}. Subsequently, Andoni, Dallant, Kozma, and Yu~\cite{adky} showed that the best product base $\gamma$ achievable in this framework satisfies $3.1818<\gamma <3.18184$.

Two recent works studied the quantum tradeoff. Caroppo et al.~\cite{cvzz}
optimized the precomputation and search parameters of the algorithm of
Ambainis et al.\ for different space budgets. Their fastest choice still
gives a time base of \(1.7274\). Dallant~\cite{dallant} used quantum
minimum finding to reduce the number of set-system tables evaluated.
For a classical construction with space and time bases \(S\) and \(T\),
the resulting quantum time base is \(\sqrt{ST}\). For the running time alone, however, the bound of~\cite{adky} on $\gamma$ above implies that \(\sqrt{ST}>1.7837\), larger than the base 1.7274 of Ambainis et al.
Thus neither the approach of Ambainis et al. nor the set-system approach improves on 1.7274 on its own, and $\Oh(1.7274^n)$ has remained the best known quantum running time for general TSP for more than seven years.

\paragraph{Our results.}
We use the quantum random-access memory (QRAM) model~\cite{cvzz}.
The number of bits or qubits per word is polynomial in the input length.
During the quantum search, reading a word of precomputed data and
reading or writing a word of quantum workspace each take one time step,
even when the address is in superposition. In this model, we prove the following.

\begin{theorem}[Running time]\label{thm:main}
There is a bounded-error quantum algorithm that finds a minimum-weight
Hamiltonian cycle in an \(n\)-city directed or undirected graph with
binary-encoded integer edge weights, or reports that none exists, in
\(O^\ast(1.70961^n)\) time. It uses \(O^\ast(1.70961^n)\) words of
classically precomputed data with coherent read access and
\(O^\ast(1.47229^n)\) additional words of writable quantum memory.
\end{theorem}
The same approach gives the same quantum upper bounds for the following problems.
\begin{corollary}\label{cor:applications}
There are bounded-error quantum algorithms with running time
\(O^\ast(1.70961^n)\) for Minimum Set Cover on an \(n\)-element
universe, Directed Feedback Arc Set on \(n\) vertices, One-Sided
Crossing Minimization with \(n\) vertices on the side whose order
is to be chosen, and Shortest Common Superstring on \(n\) input
strings. 
\end{corollary}
These algorithms improve the \(O^\ast(1.728^n)\) quantum time bounds
for Minimum Set Cover and Directed Feedback Arc Set~\cite{abi},
One-Sided Crossing Minimization~\cite{cdd}, and Shortest Common
Superstring~\cite{khadiev-superstring}. Appendix~\ref{app:applications}
defines the four problems and proves the corollary.

Our second result gives two running times under a space bound. Total
space includes the precomputed data and the writable quantum workspace.

\begin{theorem}[Space--time tradeoffs]\label{thm:tradeoff}
For the TSP instances of Theorem~\ref{thm:main}, there are
bounded-error quantum algorithms using \(O^\ast(S^n)\) total space
and \(O^\ast(T^n)\) time for
\[
  (S,T)=(1.65,1.71700)\qquad\text{and}\qquad
  (S,T)=(1.40,1.77467).
\]
\end{theorem}

\begin{table}[htbp]
\centering
\caption{Time bases for TSP at two total space budgets.
}
\label{tab:published-tsp-comparison}
\begin{tabular}{rccc}
\hline
Total space base \(S\) & Caroppo et al.~\cite{cvzz} &
Dallant~\cite{dallant} & This work \\
\hline
1.65 & 1.79001 & 1.78377 & 1.71700 \\
1.40 & 1.85870 & 1.79317 & 1.77467 \\
\hline
\end{tabular}
\end{table}

Our tradeoffs improve the previous tradeoffs of~\cite{cvzz,dallant} at every $1.40\le S \le 2$:
\begin{itemize}
    \item For \(1.40\le S<1.65\), our algorithm has time base
\(1.77467\), below the best time base on the curve of
Caroppo et al. even at the larger budget \(S=1.65\)
(approximately \(1.79001\))~\cite{cvzz}.
For \(1.65\le S\le2\), our algorithm has time base
\(1.71700\), below the minimum of their entire curve
(approximately \(1.7274\)).
\item Recall that \(\gamma\) is the infimum of \(ST\) over the choices of
set system in the classical algorithms of~\cite{anw,dk}.
For the same set system, Dallant's quantum algorithm gives a time
base of \(\sqrt{ST}\)~\cite{dallant}. Since
\(\gamma>3.1818\)~\cite{adky}, this gives
\(\sqrt{ST}\ge\sqrt{\gamma}>\sqrt{3.1818}\) for every space base \(1<S\le2\).
Our algorithm at space base \(1.40\) has time base
\(1.77467<\sqrt{3.1818}\), improving on Dallant's 
tradeoff~\cite{dallant} at every space budget
\(1.40\le S\le2\).
\end{itemize}

\paragraph*{Other related work.}
Moylett, Linden, and Montanaro~\cite{mlm-degree} used quantum
backtracking to solve weighted TSP on undirected graphs of maximum
degree three in \(O^\ast(1.110^n \operatorname{polylog}(2+L))\) time,
where \(L\) is the maximum edge cost.
Bai, Shang, and Lin~\cite{bsl} also use quantum search for TSP, and obtained a non-nested quantum search that solves undirected TSP in
\(O^\ast(1.866^n)\) time in the QRAM model.

The combination of classical precomputation and quantum search has also been adapted to other problems, such as Steiner tree~\cite{milg}, graph coloring
and the chromatic number~\cite{sm-color,gl-color}, high-dimensional set
cover~\cite{gkmv}, and treewidth~\cite{kpv}. Other applications include
optimal variable ordering for binary decision diagrams~\cite{tani-obdd},
Bayesian network structure learning~\cite{hrk},
one-sided crossing minimization~\cite{cdd}, and single-machine
scheduling~\cite{grange}.

Building on the quantum Set Cover algorithm of Ambainis et al.,
Abboud~\cite{abboud-dp} derived classical time lower bounds for
\(k\)-SUM and \(k\)-Clique under the Set Cover Conjecture, which states that Minimum Set Cover on an $n$-element universe requires $2^{n-o(n)}$-time.

\section{Extremal set systems and quantum minimum finding}\label{sec:tools}

Let \(V\) be the set of \(n\) cities. A set system on \(V\) is a family \(\mathcal F\) of subsets of \(V\). For an order \(\pi=(v_1,\ldots,v_k)\) of distinct cities, let \(P_t(\pi)=\{v_1,\ldots,v_t\}\) denote the set of the first \(t\) cities. We say that \(\mathcal F\) supports \(\pi\) if \(P_t(\pi)\in\mathcal F\) for every \(1\le t\le k\). When $k=n$, the sets $\varnothing \subset P_1(\pi)\subset\cdots \subset P_n(\pi)$ form a maximal chain of subsets of $V$, and 
every maximal chain arises from exactly one order $\pi$ in this way. Hence the number of orders of $V$ supported by 
$\mathcal F$ equals the number of maximal chains in $\mathcal F \cup \{\varnothing\}$, the quantity studied in~\cite{jlr,anw,dk,adky}.

We now state two lemmas based on previous work.

\begin{lemma}[adapted from~\cite{dk,dallant}]\label{lem:restricted}
Let \(G\) be a weighted directed graph on \(V\), and let \(\mathcal F\) be a set system on \(V\). For each \(S\in\mathcal F\) and \(u,v\in S\), let \(d_{\mathcal F}(S,u,v)\) be the minimum cost of a path from \(u\) to \(v\) that visits exactly the cities in \(S\), each once, in an order supported by \(\mathcal F\). Set \(d_{\mathcal F}(S,u,v)=+\infty\) if no such path exists.

Suppose that the members of \(\mathcal F\) are given as a list. In \(O^\ast(|\mathcal F|+1)\) time and space, we can construct a table where each entry indexed by \((S,u,v)\) stores \(d_{\mathcal F}(S,u,v)\). Each value can be retrieved in polynomial time, and a path attaining any finite value can be recovered in polynomial time. 
\end{lemma}
This lemma is slightly more general than the original statement, in the following sense. Dallant and Kozma stated their lemma for a minimum-cost tour through all cities whose visiting order is supported by \(\mathcal F\). We need the minimum costs of supported paths for every \(S\in\mathcal F\) and every pair of endpoints \(u,v\in S\). For a fixed starting city \(u\), a single run of their dynamic program already computes \(d_{\mathcal F}(S,u,v)\) for every \(S\) containing \(u\) in \(\mathcal F\) and every last city \(v\in S\). We retain these intermediate values and repeat for the \(n\) choices of \(u\), which preserves the \(O^\ast(|\mathcal F|+1)\) time and space bounds.

Our second lemma concerns set systems defined by coloring the cities. Let \(R\subseteq V\) (with \(r=|R|\)) be the set of red cities, and color the remaining cities blue. We specify a family by choosing a set \(\mathcal D\subseteq\{0,\ldots,r\}\times\{0,\ldots,n-r\}\) of allowed pairs. A city set belongs to the family if its numbers of red and blue cities form an allowed pair. Thus the family is
\[
\mathcal F_R(\mathcal D)
=\bigl\{S\subseteq V:\bigl(|S\cap R|,\,|S\setminus R|\bigr)\in\mathcal D\bigr\}.
\]
Fix a visiting order \(\pi=(v_1,\ldots,v_k)\). To determine whether \(\mathcal F_R(\mathcal D)\) supports \(\pi\), count the red and blue cities among \(v_1,\ldots,v_t\) for each \(t=1,\ldots,k\). The family supports \(\pi\) exactly when every one of these pairs belongs to \(\mathcal D\). The next lemma gives the size of the family for a fixed \(R\) and the probability of support when \(R\) is chosen uniformly among the \(r\)-element subsets of \(V\).

\begin{lemma}\label{lem:color-counts}
    For any coloring with \(r\) red cities, the family \(\mathcal F_R(\mathcal D)\) has size
\[
|\mathcal F_R(\mathcal D)|
=\sum_{(i,j)\in\mathcal D}\binom ri\binom{n-r}{j},
\]
and its members can be listed in \(O^\ast(|\mathcal F_R(\mathcal D)|+1)\) time.

Now fix a visiting order \(\pi=(v_1,\ldots,v_k)\) and choose \(R\) uniformly among the \(r\)-element subsets of \(V\). Let \(N(i,j)\) be the number of sequences of \(i\) red and \(j\) blue letters for which every nonempty prefix has its pair of color counts in \(\mathcal D\). Then
\[
\Pr\bigl[\mathcal F_R(\mathcal D)\text{ supports }\pi\bigr]
=\frac{1}{\binom nr}
  \sum_{i=0}^{k}N(i,k-i)\binom{n-k}{r-i},
\]
where a binomial coefficient is zero if its lower argument is negative or exceeds its upper argument. This probability can be computed in polynomial time.
\end{lemma}
\begin{proof}
    A set with \(i\) red cities and \(j\) blue cities is specified by choosing \(i\) cities from \(R\) and \(j\) from \(V\setminus R\). Listing these choices for each \((i,j)\in\mathcal D\) lists every member of the family exactly once. This gives both the size formula and the listing bound.

We next prove the probability in the statement. Fix the allowed pairs \(\mathcal D\), the visiting order
\(\pi=(v_1,\ldots,v_k)\), and the total number \(r\) of red cities.
For each \(i=0,\ldots,k\), we count the \(r\)-element sets \(R\subseteq V\)
for which exactly \(i\) of \(v_1,\ldots,v_k\) belong to \(R\) and
\(\mathcal F_R(\mathcal D)\) supports \(\pi\).

\begin{enumerate}
\item Color \(v_1\), then \(v_2\), and so on up to \(v_k\).
      After coloring \(v_t\), count the red and blue cities among
      \(v_1,\ldots,v_t\). For the coloring to support \(\pi\), these
      two counts must form a pair in \(\mathcal D\) at every step.
      By definition, \(N(i,k-i)\) is the number of ways to color
      these \(k\) cities that pass every check and end with exactly
      \(i\) red cities.

\item Choose \(r-i\) of the remaining \(n-k\) cities to be red and
      color the others blue. There are \(\binom{n-k}{r-i}\) choices.
      All of them preserve support for \(\pi\), since none of these
      cities occurs in \(\pi\).
\end{enumerate}
Thus, for fixed \(i\), the number of such colorings is
\[
  N(i,k-i)\binom{n-k}{r-i}.
\]
Summing over \(i\) and then dividing
by \(\binom nr\), the total number of ways to choose the \(r\) red
cities, gives the stated probability.

Starting from \(N(0,0)=1\), the dynamic program of Jabal Ameli,
Nederlof, and Wang~\cite{anw} computes all the required
\(N(i,j)\) in polynomial time by allowing a step to \((i,j)\)
exactly when \((i,j)\in\mathcal D\). Hence, the probability is also computable in polynomial time.
\end{proof}

In the running-time analysis, we will use the probability that a random
coloring satisfies two requirements at once. The family
\(\mathcal F_R(\mathcal D)\) must support the fixed visiting order
\(\pi=(v_1,\ldots,v_k)\), and each of several specified blocks outside
\(\pi\) must contain a prescribed number of red cities.

Let \(B_1,\ldots,B_b\) be a fixed partition of
\(V\setminus\{v_1,\ldots,v_k\}\). For each \(h\), let \(m_h=|B_h|\)
and require exactly \(r_h\) of these \(m_h\) cities to be red, where
\(0\le r_h\le m_h\). Since there are \(r\) red cities in total,
exactly \(i=r-\sum_{h=1}^b r_h\) of \(v_1,\ldots,v_k\) must be red.
If \(i<0\) or \(i>k\), the probability is zero.

For \(0\le i\le k\), there are \(N(i,k-i)\) ways to color
\(v_1,\ldots,v_k\) so that every prefix has an allowed pair of color
counts. Each such coloring can be completed by choosing \(r_h\) red
cities from each block \(B_h\), in \(\prod_{h=1}^b\binom{m_h}{r_h}\)
ways. Choosing \(R\) uniformly among the \(r\)-element subsets of \(V\)
therefore gives
\begin{equation}
  \Pr_R\!\left[
    \begin{gathered}
      \mathcal F_R(\mathcal D)\text{ supports }\pi
      \quad\text{and}\\
      |R\cap B_h|=r_h \text{ for every }h=1,\ldots,b
    \end{gathered}
  \right]
  =
  \frac{N(i,k-i)\displaystyle\prod_{h=1}^b\binom{m_h}{r_h}}
       {\binom nr}.
  \label{eq:support-and-block-counts}
\end{equation}

\paragraph{Quantum minimum finding.}
We will use the following lemma to minimize over choices of city sets.

\begin{lemma}[Quantum minimum finding {\cite{dh}}]
\label{lem:quantum-minimum-finding}
Let \(f(1),\ldots,f(N)\) be \(N\geq 1\) costs, and suppose that a
quantum algorithm can evaluate \(f(i)\) exactly on a superposition
of indices \(i\). For any \(0<\delta\leq 1/2\), there is a quantum
algorithm that uses \(O(\sqrt{N}\log(1/\delta))\) evaluations of \(f\)
and, with probability at least \(1-\delta\), returns an index \(i\)
minimizing \(f(i)\), together with its cost \(f(i)\).
\end{lemma}
\paragraph{Amplitude amplification.}
Suppose a quantum algorithm \(U\) returns a solution with probability at least \(p>0\) and reports failure otherwise. Given the lower bound \(p\), amplitude amplification returns a solution with probability at least \(2/3\) using
\(O(1/\sqrt p)\) calls to \(U\) and \(U^{-1}\)~\cite{bhmt}.

\section{Splitting a tour into four paths}\label{sec:four-path-search}

We now give a quantum algorithm that searches for an optimal tour by
partitioning the cities into four sets and joining four paths, one
through each set. We bound its running time and space usage in terms
of the path sizes and the coloring parameters, which we choose
explicitly in the \hyperref[sec:running-time]{next section}.

\paragraph{Tour decomposition and precomputed path costs.}
Fix a starting city and four positive integers $a,b,c,d$ whose sum is \(a+b+c+d=n\). We cut each tour into four consecutive paths of these sizes, starting from the fixed city. Write \(A,B,C,D\) with \( |A|=a,|B|=b,|C|=c,|D|=d\) for the sets of cities on the four paths, in visiting order. We choose the first path to contain the most cities: $a\ge \mathrm{max}\{b,c,d\}$.

Suppose for a moment that \(A,B,C,D\) and the two endpoints of each path have been specified. The last city of each path must be joined to the first city of the next, with the fourth path returning to the starting city. These four edges are therefore fixed. The cheapest tour consistent with the specified sets and endpoints is obtained by finding a cheapest path within each set and adding the costs of the four connecting edges. We can enumerate all choices of endpoints with a polynomial factor in the running time. It remains to search over the four city sets.

Before choosing a coloring, we build a table \(T_{\rm small}\). For every city set \(S\) with \(|S|\leq\max\{b,c,d\}\) and every \(u,v\in S\), its entry \(T_{\rm small}(S,u,v)\) stores the minimum cost of a path from \(u\) to \(v\) visiting exactly the cities in \(S\). Let \(N_{\rm small}\) be the number of city sets of size at most
\(\max\{b,c,d\}\). Then
\[
  N_{\rm small}=\sum_{j=0}^{\max\{b,c,d\}}\binom nj.
\]
By Lemma~\ref{lem:restricted}, we can build \(T_{\rm small}\) in \(O^\ast(N_{\rm small})\)
time and space.

\paragraph{Searching over partitions for a fixed coloring.}
Fix the number \(r\) of red cities and the set \(\mathcal D\) of
allowed color-count pairs. For a coloring with red set \(R\), let
\(T_R\) be the table of Lemma~\ref{lem:restricted} for \(\mathcal F_R(\mathcal D)\).
Its entry \(T_R(S,u,v)\) stores \(d_{\mathcal F_R(\mathcal D)}(S,u,v)\).
We take a missing entry to have value \(+\infty\).

For each part \(P\in\{A,B,C,D\}\), fix the first city \(s_P\)
and the last city \(t_P\) of its path. In particular, \(s_A\) is
the fixed starting city of the tour. Let \(J\) be the total weight
of the four edges that join the paths in the order \(A,B,C,D,A\).
To evaluate a candidate partition, we read the path cost for \(A\)
from \(T_R\) and the path costs for \(B,C,D\) from \(T_{\rm small}\),
using the endpoints of each part. The value of the partition is
the sum of these four costs and \(J\).

Given an integer \(K\), Figure~\ref{fig:fixed-color-search} below searches
over partitions \((A,B,C,D)\) of \(V\) to find four paths, one through
each part, that join to form a tour of total weight at most \(K\).
The algorithm groups these partitions by \(X=A\cup B\). For each
\(X\), it considers splits of \(X\) into \(A,B\) and of
\(V\setminus X\) into \(C,D\), with prescribed numbers \(r_X\)
and \(r_A\) of red cities in \(X\) and \(A\), respectively.

\begin{lemma}[Threshold search for a fixed coloring]
\label{lem:fixed-color-search}
Fix an input to the algorithm in Figure~\ref{fig:fixed-color-search}. Suppose some partition \((A,B,C,D)\) of the prescribed sizes and
red-city counts satisfies
\[
  T_R(A,s_A,t_A)
  + \sum_{P\in\{B,C,D\}} T_{\rm small}(P,s_P,t_P)
  + J \leq K.
\]
Then Figure~\ref{fig:fixed-color-search} returns a tour of weight
at most \(K\) with probability at least \(2/3\).
Its running time, including the construction of \(T_R\), is
\[
  O^\ast\!\left(
    |\mathcal F_R(\mathcal D)|
    +\sqrt{N_X}\bigl(\sqrt{N_A}+\sqrt{N_C}\bigr)
  \right),
\]
where \(N_X\) counts the choices of \(X\), and \(N_A,N_C\) count
the choices of \(A,C\), respectively, for each fixed \(X\):
\[
  N_X=\binom{r}{r_X}\binom{n-r}{a+b-r_X},
  \qquad
  N_A=\binom{r_X}{r_A}\binom{a+b-r_X}{a-r_A},
  \qquad
  N_C=\binom{c+d}{c}.
\]
\end{lemma}

\begin{figure}[!htbp]
\centering
\begin{minipage}{0.96\linewidth}
\small
\hrule
\smallskip
\textbf{Input.} The cities \(V\), edge weights \(w\), precomputed
table \(T_{\rm small}\), path sizes \(a,b,c,d\), endpoint pairs
\((s_P,t_P)\) for \(P\in\{A,B,C,D\}\), coloring \(R\), allowed
color-count pairs \(\mathcal D\), red-city counts \(r_X,r_A\),
and threshold \(K\).

\smallskip
\textbf{Output.} A verified tour of weight at most \(K\), or failure.

\begin{tabbing}
\hspace{1.4em}\=\kill
\textbf{if} any of \(N_X,N_A,N_C\) is zero, \textbf{return} failure.\\
Build \(T_R\) by applying Lemma~\ref{lem:restricted} to
\(\mathcal F_R(\mathcal D)\).\\[0.5ex]
\textbf{Subroutine} \textsc{Check}\\
\> Prepare a uniform superposition of the \(N_X\) choices of \(X\).\\
\> \(Y \gets V\setminus X\).\\[0.4ex]
\> \(\displaystyle
   A \gets \operatorname*{arg\,min}_{\substack{S\subseteq X,\ |S|=a\\
                                               |S\cap R|=r_A}}
   \bigl[T_R(S,s_A,t_A)+T_{\rm small}(X\setminus S,s_B,t_B)\bigr]
   \).\\[0.6ex]
\> \(\displaystyle
   C \gets \operatorname*{arg\,min}_{\substack{S\subseteq Y,\ |S|=c}}
   \bigl[T_{\rm small}(S,s_C,t_C)+T_{\rm small}(Y\setminus S,s_D,t_D)\bigr]
   \).\\[0.6ex]
\> \(B\gets X\setminus A,\qquad D\gets Y\setminus C\).\\
\> Recover the path for \(A\) from \(T_R\) and the paths for \(B,C,D\) from \(T_{\rm small}\).\\
\> \textbf{if} any path is missing, \textbf{reject}.\\
\> Join the four paths using the prescribed edges to obtain \(\tau\).\\
\> \textbf{accept} and retain \(\tau\) if \(\tau\) is a tour of weight at most \(K\).\\
\> \textbf{otherwise reject}.\\[0.5ex]
Apply amplitude amplification using \(O(\sqrt{N_X})\) calls
to \textsc{Check} and its inverse.\\
\textbf{return} the verified tour if found, and failure otherwise.
\end{tabbing}
\hrule
\end{minipage}
\caption{Threshold test for a fixed coloring and fixed path endpoints.
Each minimization uses Lemma~\ref{lem:quantum-minimum-finding}
with error probability at most \(1/10\).}
\label{fig:fixed-color-search}
\end{figure}

\begin{proof}
Fix a partition \((A_0,B_0,C_0,D_0)\) satisfying the hypothesis
of the lemma, and put \(X_0=A_0\cup B_0\). We show that choosing \(X=X_0\) suffices
for \textsc{Check} to find a tour of weight at most \(K\), provided
that both minimum searches succeed.

To see this, consider what remains to be chosen once \(X\) is fixed.
Within \(X\), choosing \(A\) determines \(B=X\setminus A\).
Independently, within \(Y=V\setminus X\), choosing \(C\) determines
\(D=Y\setminus C\). Since the prescribed endpoints already fix
the cost \(J\) of joining the paths, the minimum partition value
for this \(X\) is \(J+\operatorname{Cost}(X)\), where
\[
\begin{aligned}
\operatorname{Cost}(X)
={}&\min_{\substack{A\subseteq X,\ |A|=a\\ |A\cap R|=r_A}}
    \bigl[T_R(A,s_A,t_A)+T_{\rm small}(X\setminus A,s_B,t_B)\bigr]\\
 &+\min_{\substack{C\subseteq Y\\ |C|=c}}
    \bigl[T_{\rm small}(C,s_C,t_C)+T_{\rm small}(Y\setminus C,s_D,t_D)\bigr].
\end{aligned}
\]
In particular, when \(X=X_0\), the two minima can use \(A_0\) and
\(C_0\), so \(J+\operatorname{Cost}(X_0)\leq K\).

It remains to bound the probability of finding such a tour.
Choosing \(r_X\) red cities and \(a+b-r_X\) blue cities uniformly
gives \(X=X_0\) with probability \(1/N_X\). Conditional on this
choice, Lemma~\ref{lem:quantum-minimum-finding} with error probability
\(1/10\) for each search in \textsc{Check} gives probability at least \(4/5\) that both return minimizers. Their paths then pass verification by the bound
above. Thus \textsc{Check} succeeds with probability at least
\(4/(5N_X)\), and amplitude amplification raises this probability
to at least \(2/3\) using \(O(\sqrt{N_X})\) calls to \textsc{Check}
and its inverse.

To bound the cost of these calls, observe that \(A\) must contain
\(r_A\) of the \(r_X\) red cities in \(X\) and \(a-r_A\) of its
blue cities, giving \(N_A\) choices. For \(C\), any \(c\) of the
\(c+d\) cities outside \(X\) may be chosen, giving \(N_C\) choices.
The two minimum searches therefore take
\(O^\ast(\sqrt{N_A}+\sqrt{N_C})\) time per call.
Multiplying by the number of calls and adding the cost
\(O^\ast(|\mathcal F_R(\mathcal D)|)\) of building \(T_R\)
by Lemma~\ref{lem:restricted} gives the stated running time.
\end{proof}

\paragraph{Amplification over colorings and total complexity.}
We now use Lemma~\ref{lem:fixed-color-search} to search over
colorings. We build \(T_{\rm small}\) once. For each choice of
path endpoints, \textsc{Trial} in Figure~\ref{fig:amplify-colorings}
prepares a uniform superposition of the \(r\)-element sets
\(R\subseteq V\) and runs Figure~\ref{fig:fixed-color-search}
coherently on \(R\). To bound the success probability of
\textsc{Trial}, we identify the colorings for which the partition
of an optimal tour satisfies the hypothesis of
Lemma~\ref{lem:fixed-color-search}. Fix an optimal tour \(\tau^\star\)
of weight \(W^\star\). Starting at the prescribed city, divide
\(\tau^\star\) into four consecutive paths of \(a,b,c,d\) cities,
with city sets \(A^\star,B^\star,C^\star,D^\star\), respectively.
Put \(X^\star=A^\star\cup B^\star\), and let \(\pi_A^\star\) be the
order in which \(\tau^\star\) visits \(A^\star\). Let \(p>0\) be
a known lower bound on the probability that a uniformly chosen
\(r\)-element set \(R\subseteq V\) satisfies both conditions below.
\begin{enumerate}
\item \(|X^\star\cap R|=r_X\) and \(|A^\star\cap R|=r_A\).
\item \(\mathcal F_R(\mathcal D)\) supports \(\pi_A^\star\).
\end{enumerate}

\begin{figure}[!htbp]
\centering
\begin{minipage}{0.96\linewidth}
\small
\hrule
\smallskip
\textbf{Input.}
The weighted graph and starting city, the precomputed table
\(T_{\rm small}\), the path sizes \(a,b,c,d\), the coloring
parameters \(r,\mathcal D,r_X,r_A\), the threshold \(K\),
and the lower bound \(p\).

\smallskip
\textbf{Output.}
A verified tour of weight at most \(K\), or failure.

\begin{tabbing}
\hspace{1.4em}\=\hspace{1.4em}\=\kill
\textbf{for each} choice of the four endpoint pairs \((s_P,t_P)\), $P\in\{A,B,C,D\}$
with \(s_A\) fixed to the starting city \textbf{do}\\
\> \textbf{Subroutine} \textsc{Trial} for the current endpoints\\
\>\> Prepare a uniform superposition of the \(r\)-element sets \(R\subseteq V\).\\
\>\> Run Figure~\ref{fig:fixed-color-search} coherently on \(R\).\\
\> Amplify the probability that \textsc{Trial} produces a verified tour,\\
\> using \(O(1/\sqrt p)\) calls to \textsc{Trial} and its inverse.\\
\> \textbf{if} a verified tour is obtained, \textbf{return} the tour.\\
\textbf{return} failure.
\end{tabbing}
\hrule
\end{minipage}
\caption{A protocol to decide whether there exists a tour of weight at most $K$.}
\label{fig:amplify-colorings}
\end{figure}

Suppose \(K\geq W^\star\) and consider the iteration in which
\(s_P,t_P\) are the first and last cities of \(P^\star\) along
\(\tau^\star\), for every \(P\in\{A,B,C,D\}\).
For any \(R\) satisfying both conditions, the searches can choose
\(X^\star,A^\star,C^\star\), and \(T_R(A^\star,s_A,t_A)\) is at most
the weight of the path through \(A^\star\) in \(\tau^\star\).
Hence \(J+\operatorname{Cost}(X^\star)\leq W^\star\leq K\).
Lemma~\ref{lem:fixed-color-search} therefore guarantees that
Figure~\ref{fig:fixed-color-search} returns a verified tour with
probability at least \(2/3\) for every coloring satisfying both
conditions. Since these colorings have probability at least \(p\),
one call to \textsc{Trial} with the endpoints of \(\tau^\star\)
succeeds with probability at least \(2p/3\).
Figure~\ref{fig:amplify-colorings} thus uses \(O(1/\sqrt p)\)
calls to \textsc{Trial} and its inverse. Multiplying the running
time in Lemma~\ref{lem:fixed-color-search} by this factor and
adding the cost $\Oh(N_{\rm small})$ of \(T_{\rm small}\) gives
\begin{equation}
  O^\ast\!\left(
    N_{\rm small}
    +\frac{|\mathcal F_R(\mathcal D)|
      +\sqrt{N_X}\bigl(\sqrt{N_A}+\sqrt{N_C}\bigr)}
      {\sqrt p}
  \right).
  \label{eq:four-path-time}
\end{equation}
By Lemma~\ref{lem:color-counts}, \(|\mathcal F_R(\mathcal D)|\)
is the same for every \(R\) of size \(r\). The algorithm uses
\(O^\ast(N_{\rm small})\) words of precomputed data and
\(O^\ast(|\mathcal F_R(\mathcal D)|)\) additional words of
writable quantum memory.

Finally, if \(M\) bounds the absolute value of every edge weight,
the test at \(K=nM\) decides whether a tour exists. On a feasible
instance, binary search gives the optimum, and a final test at that value returns an optimal tour. Enumerating the endpoint choices, testing the thresholds, and
repeating the tests to keep their total error probability at most
\(1/3\) cost only a polynomial factor. Hence there exists a bounded-error quantum algorithm for TSP with running time~\eqref{eq:four-path-time}.

\section{Proof of the running time bound}\label{sec:running-time}

We prove the \(O^\ast(1.70961^n)\) running-time bound of
Theorem~\ref{thm:main} by evaluating~\eqref{eq:four-path-time}.
The algorithm builds \(T_{\rm small}\) once. Each call to
\textsc{Trial} then builds \(T_R\) and searches for a partition under
the chosen coloring \(R\). We first choose the parameters and bound
the time for precomputation and for one call. It remains to bound
the number \(O(1/\sqrt p)\) of calls to \textsc{Trial} and its inverse.
For this we lower-bound \(p\) by counting color sequences along
the first path of a fixed optimal tour.

\paragraph{Precomputation and the cost of one trial.}
The size of the largest of the three shorter paths determines the
cost of building \(T_{\rm small}\). Write \(n=100000k\), padding the
instance if necessary,\footnote{To reach the next multiple of
\(100000\), replace one city by a directed path of zero-weight edges.
Attach its former incoming edges to the first city of the path and
its former outgoing edges to the last. Every tour traverses the
entire path, so the optimum is preserved. This adds fewer than
\(100000\) cities. For an undirected input, first replace each edge
by two oppositely directed edges of the same weight.}
and choose the four path sizes to be
\[
 (a,b,c,d)=k(35051,22752,22752,19445).
\]
The three shorter paths have at most \(22752k\) cities each, so
the number of city sets for which we precompute path costs is
\[
 N_{\rm small}=\sum_{j=0}^{22752k}\binom{n}{j}
              =O^\ast(1.709605^n)
\]
by Stirling's formula.

To bound the cost of building \(T_R\) within a trial, we limit the
number of city sets included for each pair of color counts.
Choose \(60606k\) red cities uniformly at random and color the
remaining \(39394k\) cities blue. Include a pair \((i,j)\) in
\(\mathcal D\) precisely when
\[
 \binom{60606k}{i}\binom{39394k}{j}\leq 2^{0.55805810n}.
\]
The left-hand side counts the city sets with exactly \(i\) red cities
and \(j\) blue cities. Since there are at most \((n+1)^2\) such pairs, we have
\[
  |\mathcal F_R(\mathcal D)|\le (n+1)^2\,2^{0.55805810n}.
\]

To bound the search time within the same trial, set \(r_X=20873k\)
and \(r_A=4952k\) in Figure~\ref{fig:fixed-color-search}. A candidate \(X\) therefore
contains \(20873k\) red cities and \(36930k\) blue cities. Within
such an \(X\), a candidate \(A\) contains \(4952k\) red cities and
\(30099k\) blue cities. For \(C\), we consider all \(22752k\)-element
subsets of the \(42197k\) cities outside \(X\). The numbers of candidates are
\begin{align*}
 N_X &= \binom{60606k}{20873k}\binom{39394k}{36930k},\\
 N_A &= \binom{20873k}{4952k}\binom{36930k}{30099k},\\
 N_C &= \binom{42197k}{22752k}.
\end{align*}
By Stirling's formula, both \(\sqrt{N_XN_A}\) and \(\sqrt{N_XN_C}\)
are \(O^\ast(2^{0.55805810n})\). Together with the bound on
\(|\mathcal F_R(\mathcal D)|\), this gives \(O^\ast(2^{0.55805810n})\)
time per call to \textsc{Trial}.

\paragraph{The coloring probability \(p\).}
To bound the number of trials in~\eqref{eq:four-path-time}, we need
a lower bound \(p\) on the probability that the fixed optimal tour
in Section~\ref{sec:four-path-search} satisfies the coloring requirements. Its sets \(A^\star\)
and \(X^\star=A^\star\cup B^\star\) must have their required red
counts, and \(\mathcal F_R(\mathcal D)\) must support the visiting
order \(\pi_A^\star\) on the \(35051k\) cities of \(A^\star\).
The required red counts leave
\((20873-4952)k=15921k\) red cities in \(B^\star\) and
\((60606-20873)k=39733k\) outside \(X^\star\).
By~\eqref{eq:support-and-block-counts}, we may therefore take
\begin{equation}\label{eq:fixed-word-success}
  p =
  \frac{N(4952k,30099k)
        \binom{22752k}{15921k}\binom{42197k}{39733k}}
       {\binom{100000k}{60606k}}.
\end{equation}
Thus a lower bound on
\(N(4952k,30099k)\) gives the required lower bound on \(p\).
Here, the sequences counted by \(N(4952k,30099k)\) are the color sequences
along \(\pi_A^\star\) satisfying both of the following conditions.
\begin{enumerate}
    \item The full sequence must
contain \(4952k\) red letters and \(30099k\) blue letters.
\item For every
prefix, the numbers \(i\) of red letters and \(b\) of blue letters
must satisfy \((i,b)\in\mathcal D\), equivalently,
\begin{equation}\label{eq:fixed-word-prefix}
  \binom{60606k}{i}\binom{39394k}{b}
  \le 2^{0.55805810n}=2^{55805.81k}.
\end{equation}
\end{enumerate} 
To lower-bound \(N(4952k,30099k)\), we divide \(\pi_A^\star\) into
four consecutive intervals and fix a red count for each interval.
The next lemma gives conditions under which at least
half of these choices of red positions satisfy~\eqref{eq:fixed-word-prefix}
at every prefix.

\begin{lemma}\label{lem:space-prefix-count}
Let \(r\) of the \(n\) cities be red. For a fixed \(\ell>0\), call a
pair \((i,b)\) allowed if \(\binom ri\binom{n-r}{b}\le 2^{\ell n}\).
Divide the positions of the first path into a fixed number \(m\) of
nonempty consecutive intervals \(A_1,\ldots,A_m\). Fix a number
\(r_{A_j}\) of red positions in each \(A_j\), with \(0\le r_{A_j}\le |A_j|\).
Let \(I_j\) and \(J_j\) be the total numbers of red and blue positions,
respectively, in the first \(j\) intervals. 

For each interval \(A_j\), choose positive constants \(u_j,v_j\)
independent of \(n\) and define
\begin{equation}\label{eq:space-binomial-bound}
 L_j(i,b)=r\log_2(1+u_j)+(n-r)\log_2(1+v_j)
                 -i\log_2 u_j-b\log_2 v_j .
\end{equation}
Suppose there is a constant \(\eta<\ell\), independent of \(n\), such that
\[
 L_j(I_{j-1},J_{j-1})\le\eta n
 \quad\text{and}\quad L_j(I_j,J_j)\le\eta n
 \qquad (1\le j\le m).
\]
For all sufficiently large \(n\), at least half of the color sequences
with exactly \(r_{A_j}\) red positions in each \(A_j\) have an allowed
pair of counts at every prefix. In particular,
\[
  N(I_m,J_m)\ \ge\ \tfrac12\prod_{j=1}^{m}\binom{|A_j|}{r_{A_j}} .
\]
\end{lemma}

\begin{proof}
There are \(M=\prod_{j=1}^m\binom{|A_j|}{r_{A_j}}\) color sequences
with the chosen red counts in the intervals. All \(M\) sequences have
\(I_m\) red and \(J_m\) blue positions in total. We show that at most
\(M/2\) have a prefix whose pair of counts is outside the allowed set.

Fix \(j\) and \(t\) with \(1\le j\le m\) and \(1\le t\le |A_j|\).
Let \(P_{j,t}\) be the prefix consisting of all positions in
\(A_1,\ldots,A_{j-1}\) followed by the first \(t\) positions of \(A_j\).
Put \(z=t/|A_j|\). Among the \(M\) sequences, each
position of \(A_j\) is red in a fraction \(r_{A_j}/|A_j|\) of the
sequences. The first \(t\) positions of \(A_j\) therefore contain
\(z r_{A_j}\) red positions on average. Including the previous
intervals, the average red and blue counts in \(P_{j,t}\) are
\[
\begin{aligned}
(\bar i,\bar b)
&= \underbrace{(I_{j-1},J_{j-1})}_{\text{counts before }A_j}
   + \underbrace{z\bigl(r_{A_j},\,|A_j|-r_{A_j}\bigr)}
                _{\text{average counts in the first }t\text{ positions of }A_j} \\
&= (I_{j-1},J_{j-1})
   + z\bigl(I_j-I_{j-1},\,J_j-J_{j-1}\bigr) \\
&= (1-z)(I_{j-1},J_{j-1}) + z(I_j,J_j).
\end{aligned}
\]
Since \(L_j\) is affine,
\[
 L_j(\bar i,\bar b)
 = (1-z)L_j(I_{j-1},J_{j-1}) + zL_j(I_j,J_j).
\]
By the assumption, both \(L_j(I_{j-1},J_{j-1})\) and \(L_j(I_j,J_j)\)
are at most \(\eta n\). It follows that
\(L_j(\bar i,\bar b)\le \eta n\).

Consider a color sequence with \(i\) red and \(b\) blue positions in
\(P_{j,t}\). If \((i,b)\) is outside the allowed set, the definition
of an allowed pair and the binomial theorem give
\[
  2^{\ell n}
  < \binom ri\binom{n-r}{b}
  \le \frac{(1+u_j)^r(1+v_j)^{n-r}}{u_j^i v_j^b}
  = 2^{L_j(i,b)}.
\]
Thus \(L_j(i,b)>\ell n\), whereas the average counts in \(P_{j,t}\)
satisfy \(L_j(\bar i,\bar b)\le\eta n\).
Write \(d=i-\bar i\). Since \(i+b=\bar i+\bar b=|P_{j,t}|\),
we have \(b-\bar b=-d\). Substituting in the definition of \(L_j\) gives
\[
  (\ell-\eta)n
  < L_j(i,b)-L_j(\bar i,\bar b)
  = d\log_2(v_j/u_j).
\]
For sufficiently large \(n\),
\((\ell-\eta)n>n^{2/3}|\log_2(v_j/u_j)|\).
Hence only sequences with \(|d|>n^{2/3}\) can violate the prefix
condition on \(P_{j,t}\).

Let \(B_{j,t}\) be the number of the \(M\) sequences that violate the
prefix condition at the \(t\)-th position of \(A_j\). Choose one of the \(M\) sequences uniformly. Since a violation at the
\(t\)-th position of \(A_j\) requires \(|d|>n^{2/3}\), we have
\[
  B_{j,t}/M \le \Pr(|d|>n^{2/3}) \le O\!\left(\exp\!\left(-n^{4/3}/t\right)\right)
\]
Here the last inequality used Hoeffding's inequality~\cite{hoeffding} to
the red count \(i-I_{j-1}\) in the first \(t\) positions of \(A_j\), which has mean
\(t r_{A_j}/|A_j|\). Summing over $j$ and $t$ gives
\[
\begin{aligned}
\sum_{j=1}^{m}\sum_{t=1}^{|A_j|} B_{j,t}
&\le M\sum_{j=1}^{m}\sum_{t=1}^{|A_j|}O\left(e^{-n^{4/3}/t}\right)\\
&\le M\sum_{j=1}^{m} \sum_{t=1}^{|A_j|}O\left(e^{-n^{1/3}}\right)
       \\
&=M\cdot O\left(e^{-n^{1/3}}\right)\sum_{j=1}^{m}|A_j|\\
&\le M\cdot O\left(n e^{-n^{1/3}}\right)\\
&\le M/2.
\end{aligned}
\]
for all sufficiently large \(n\). At least \(M/2\) of the \(M\)
sequences therefore have allowed counts at every prefix. This proves the lemma.
\end{proof}

\paragraph{Interval counts for \(N(4952k,30099k)\).}
For the first path of the fixed tour, we use the four intervals
in Table~\ref{tab:word-type-bounds}. Their lengths sum to \(35051k\)
and their red counts to \(4952k\). To apply
Lemma~\ref{lem:space-prefix-count}, it suffices to check that the two
upper bounds on \(L_j/k\) in each row are strictly below \(55805.81\),
the exponent in~\eqref{eq:fixed-word-prefix} divided by \(k\).

\begin{table}[!htbp]
\centering
\caption{Intervals on the first path. The last two columns give upper
bounds at the prefixes ending just before and at the end of \(A_j\).}
\label{tab:word-type-bounds}
\begin{tabular}{crrcccc}
\hline
Interval & \(|A_j|/k\) & \(r_{A_j}/k\) & \(u_j\) & \(v_j\) &
 \(L_j(I_{j-1},J_{j-1})/k\) & \(L_j(I_j,J_j)/k\)\\
\hline
\(A_1\) & 27800 & 3201 & \(1/19\) & \(22/15\) & 55798.1 & 55803.7\\
\(A_2\) &  2700 &  495 & \(1/17\) & \(17/9\)  & 55804.3 & 55804.4\\
\(A_3\) &  2700 &  672 & \(1/14\) & \(12/5\)  & 55801.6 & 55798.7\\
\(A_4\) &  1851 &  584 & \(3/35\) & \(31/10\) & 55802.0 & 55803.8\\
\hline
\end{tabular}
\end{table}

The largest entry in the last two columns of Table~\ref{tab:word-type-bounds}
is \(55804.4\), so Lemma~\ref{lem:space-prefix-count} applies with
\(\eta=55804.4k/n=0.558044<0.55805810=\ell\). Hence
\[
 N(4952k,30099k)\ \ge\ \tfrac12
 \binom{27800k}{3201k}\binom{2700k}{495k}
 \binom{2700k}{672k}\binom{1851k}{584k}
\]
for all sufficiently large \(k\).

\paragraph{Total running time.}
Substituting the lower bound on \(N(4952k,30099k)\)
in~\eqref{eq:fixed-word-success} and applying Stirling's formula gives
\(p^{-1/2}=O^\ast(2^{0.215476510n})\). Multiplying by the cost
\(O^\ast(2^{0.55805810n})\) per trial bounds the time spent searching
over colorings. Finally, adding the cost of building \(T_{\rm small}\)
in~\eqref{eq:four-path-time} gives
\[
  O^\ast\!\left(
    \underbrace{1.709605^n}_{\text{building }T_{\rm small}}
    +\underbrace{1.709453^n}_{\text{amplification over colorings}}
  \right)
  = O^\ast(1.709605^n).
\]
This proves the running-time bound of Theorem~\ref{thm:main}.

\section{Time under a space bound}\label{sec:space-bound}

Section~\ref{sec:running-time} precomputes path costs for every set of at most \(0.22752n\)
cities. To use less space, we lower this limit by shortening \(B,C,D\),
which leaves more cities for \(A\). We use Lemma~\ref{lem:space-prefix-count}
to estimate how often a random coloring supports the visiting order on
the longer path.

We give two choices of path sizes, one for the four-path
algorithm at space base \(1.65\) and one for a six-path algorithm at
space base \(1.40\). Both choices use the same lower bound on the probability \(p\) in
Section~\ref{sec:four-path-search}. Fix an optimal tour, let \(A,B\) be the city sets of its
first two paths, and put \(X=A\cup B\) and \(Y=V\setminus X\).
Divide the visiting order on \(A\) into consecutive intervals, and
write \(A_1,\ldots,A_m\) for the corresponding city sets.
For each \(Z\in\mathcal P=\{A_1,\ldots,A_m,B,Y\}\), choose a target
red count \(r_Z\) between \(0\) and \(|Z|\). Set \(r=\sum_{Z\in\mathcal P}r_Z\) and choose
the \(r\) red cities uniformly among all \(r\)-subsets of \(V\).
For each space bound, we choose a constant \(\ell>0\) and use the set
of allowed pairs
\[
  \mathcal D=\left\{(i,b)\in\{0,\ldots,r\}\times\{0,\ldots,n-r\}:
  \binom ri\binom{n-r}{b}\le 2^{\ell n}\right\}.
\]
For the search, set
\[
  r_A=\sum_{j=1}^{m}r_{A_j},\qquad r_X=r_A+r_B.
\]
For \(p\) in~\eqref{eq:four-path-time}, take the probability that
\(|A\cap R|=r_A\), \(|X\cap R|=r_X\), and \(\mathcal F_R(\mathcal D)\)
supports the visiting order on \(A\).
Applying~\eqref{eq:support-and-block-counts} with \(B_1=B\) and \(B_2=Y\),
and using Lemma~\ref{lem:space-prefix-count} to bound
\(N(r_A,|A|-r_A)\) from below, gives
\begin{equation}\label{eq:space-coloring-probability}
  p\ge\frac{1}{2\binom nr}\prod_{Z\in\mathcal P}\binom{|Z|}{r_Z}.
\end{equation}

\subsection{Four paths at space base 1.65}
\label{sec:space-four-example}

We now choose parameters giving time \(O^\ast(1.71700^n)\) within
total space \(O^\ast(1.65^n)\). Write \(n=100000k\), using the padding
of Section~\ref{sec:running-time}. In Table~\ref{tab:space65-counts}, the red and blue
columns determine the part sizes and red counts. Take \(|C|=|B|\),
so the row for \(Y=C\cup D\) also determines \(|D|\).
For the writable table choose \(\ell=0.530633\). The last three
columns verify the inequalities of Lemma~\ref{lem:space-prefix-count}
for this value of \(\ell\).

\begin{table}[htbp]
\centering
\caption{Parameters for the algorithm at space base \(1.65\).
For each \(A_j\), the last column bounds the larger of the two
values of \(L_j/n\) in Lemma~\ref{lem:space-prefix-count}.}
\label{tab:space65-counts}
\begin{tabular}{crrccc}
\hline
Part & Red & Blue & \(u_j\) & \(v_j\) & Upper bound \\
\hline
\(A_1\) & 2224  & 26247 & 0.035  & 1.3285 & 0.53063259 \\
\(A_2\) & 1470  & 5249  & 0.0511 & 2.3    & 0.53062482 \\
\(A_3\) & 1854  & 2887  & 0.095  & 4.534  & 0.53063002 \\
\(B\)   & 15164 & 4862  & & & \\
\(Y\)   & 38169 & 1874  & & & \\
\hline
Total   & 58881 & 41119 & & & \\
\hline
\end{tabular}
\end{table}

\paragraph{Space.}
The red and blue counts of Table~\ref{tab:space65-counts} gives
\((|A|,|B|,|C|,|D|)=k(39931,20026,20026,20017)\).
Thus \(T_{\rm small}\) stores path costs for
\(N_{\rm small}=\sum_{s=0}^{20026k}\binom ns=O^\ast(1.64998^n)\)
city sets. For \(T_R\), each pair of
red and blue counts admitted by \(\mathcal D\) contributes at most
\(2^{\ell n} = 2^{0.530633n}\) city sets. There are at most \((n+1)^2\) pairs,
so the writable space for \(T_R\) is \(O^\ast(1.44457^n)\).
The total space is \(O^\ast(1.65^n)\).

\paragraph{Time.}
Applying Stirling's formula to~\eqref{eq:space-coloring-probability}
with the counts in Table~\ref{tab:space65-counts} gives
\(p^{-1/2}=O^\ast(2^{0.249255n})\).

Table~\ref{tab:space65-counts} give
\(r_A=5548k\) and \(r_X=5548k+15164k=20712k\). For \(C\), it chooses any \(20026k\) cities
of \(Y\). The numbers of candidates in~\eqref{eq:four-path-time}
are therefore
\[
\begin{aligned}
 N_X&=\binom{58881k}{20712k}\binom{41119k}{39245k},&
 N_A&=\binom{20712k}{5548k}\binom{39245k}{34383k},&
 N_C&=\binom{40043k}{20026k}.
\end{aligned}
\]
Substituting these numbers and the bound on \(p^{-1/2}\)
in~\eqref{eq:four-path-time} gives
\[
 O^\ast\!\left(
 \underbrace{1.64998^n}_{N_{\rm small}}
 +\underbrace{1.71700^n}_{|\mathcal F_R(\mathcal D)|/\sqrt p}
 +\underbrace{1.70827^n}_{\sqrt{N_XN_A/p}}
 +\underbrace{1.71699^n}_{\sqrt{N_XN_C/p}}
 \right)=O^\ast(1.71700^n).
\]

\subsection{Six paths at space base 1.40}
\label{sec:space-six-search}

At space base \(1.40\), we precompute path costs for sets of at most
\(10525k\) cities, where \(n=100000k\). With decompositions to four paths, at least \(68425k\) cities would
remain for \(A\), which is too large for our purpose. We therefore use six consecutive paths \(A,B,C,D,E,F\).
The extra two paths allow us to take \(|A|=59660k\) while keeping
each of the other five paths small enough for \(T_{\rm small}\).
We first bound the cost of dividing \(Y=V\setminus(A\cup B)\) among
\(C,D,E,F\).

\paragraph{The search on \(Y\).}
As in Section~\ref{sec:four-path-search}, fix the endpoints of all six paths before searching
over \(X=A\cup B\). These endpoints determine the costs of the six
connecting edges. For each choice of \(X\), divide its complement
\(Y\) into four paths with \(|C|=|E|=|B|\) and
\(|D|=|F|=|Y|/2-|B|\).
For each choice of \(Y_1=C\cup D\), write \(Y_2=Y\setminus Y_1=E\cup F\).
The minimum total cost of the four paths on \(Y\), with their endpoints
fixed, is
\[
\min_{\substack{Y_1\subseteq Y\\ |Y_1|=|Y|/2}}
\left(
\begin{aligned}
&\min_{\substack{C\subseteq Y_1\\ |C|=|B|}}
 \bigl[T_{\rm small}(C,s_C,t_C)
      +T_{\rm small}(Y_1\setminus C,s_D,t_D)\bigr] \\
{}+{}&
 \min_{\substack{E\subseteq Y_2\\ |E|=|B|}}
 \bigl[T_{\rm small}(E,s_E,t_E)
      +T_{\rm small}(Y_2\setminus E,s_F,t_F)\bigr]
\end{aligned}
\right).
\]
For fixed \(Y_1\), the two inner minima can be found independently,
each by quantum minimum finding over \(\binom{|Y|/2}{|B|}\) choices.
The outer minimum ranges over \(\binom{|Y|}{|Y|/2}\) choices of \(Y_1\).
The resulting search takes \(O^\ast(\sqrt{N_Y})\) time, where
\begin{equation}\label{eq:space-six-y}
 N_Y=\binom{|Y|}{|Y|/2}\binom{|Y|/2}{|B|}.
\end{equation}
Thus \(N_Y\) replaces \(N_C\) in~\eqref{eq:four-path-time}.

\paragraph{Space.}
We first show that the six-path algorithm uses \(O^\ast(1.40^n)\)
total space. The red and blue counts in Table~\ref{tab:space40-counts} below
determine the path sizes used to bound the space for \(T_{\rm small}\).
For the writable table \(T_R\), we choose \(\ell=0.4854\).
More precisely, with \(52483k\) red cities and \(47517k\) blue cities
(Table~\ref{tab:space40-counts}), the allowed pairs are given by
\[
  (i,b)\in\mathcal D
  \quad\Longleftrightarrow\quad
  \underbrace{\binom{52483k}{i}\binom{47517k}{b}}
             _{\text{number of city sets with }i\text{ red and }b\text{ blue cities}}
  \le 2^{0.4854n} < 1.40^n .
\]
Table~\ref{tab:space40-counts} gives \(|Y|=29815k\). Since
\(|B|=|C|=|E|=10525k\), we have
\[
  |D|=|F|=\frac{|Y|}{2}-|B|=\frac{8765k}{2}<10525k.
\]
To make these sizes integral, we pad to a multiple of \(200000\)
as in Section~\ref{sec:running-time}, so that \(k\) is even.
\begin{table}[htbp]
\centering
\caption{Parameters for the algorithm at space base \(1.40\).
Red and blue counts are in units of \(k\). For each \(A_j\), the last
column bounds the larger of the two values of \(L_j/n\) in
Lemma~\ref{lem:space-prefix-count}.}
\label{tab:space40-counts}
\begin{tabular}{crrccc}
\hline
Part & Red & Blue & \(u_j\) & \(v_j\) & Upper bound on \(L_j/n\) \\
\hline
\(A_1\) & 118   & 29090 & 0.002 & 1.0255 & 0.48537999 \\
\(A_2\) & 1612  & 8487  & 0.015 & 2.22   & 0.48538689 \\
\(A_3\) & 4692  & 6498  & 0.084 & 5.98   & 0.48537003 \\
\(A_4\) & 6435  & 2728  & 0.25  & 26.3   & 0.48535912 \\
\(B\)   & 10074 & 451   & & & \\
\(Y\)   & 29552 & 263   & & & \\
\hline
Total   & 52483 & 47517 & & & \\
\hline
\end{tabular}
\end{table}

The precomputed table \(T_{\rm small}\) stores path costs for
\(N_{\rm small}=\sum_{s=0}^{10525k}\binom ns=O^\ast(1.40^n)\)
city sets. As in the
four-path construction, \(|\mathcal F_R(\mathcal D)|\le(n+1)^2 2^{\ell n}\).
Hence the writable space is \(O^\ast(1.39998^n)\), and the two
tables together fit in \(O^\ast(1.40^n)\) total space.

\paragraph{Time.}
From Table~\ref{tab:space40-counts}, we may take
\(\eta=0.48538689<\ell\) in Lemma~\ref{lem:space-prefix-count}
and set \(r=52483k\), \(r_A=12857k\), and \(r_X=22931k\). Now applying Stirling's
formula in~\eqref{eq:space-coloring-probability} gives \(p^{-1/2}=O^\ast(2^{0.3421501n})\).
The searches for \(X\) and \(A\) have
\[
 N_X=\binom{52483k}{22931k}\binom{47517k}{47254k},
 \qquad
 N_A=\binom{22931k}{12857k}\binom{47254k}{46803k}
\]
candidates. Using \(|Y|=29815k\) and \(|B|=10525k\)
in~\eqref{eq:space-six-y}, the four terms of~\eqref{eq:four-path-time}
are bounded by
\[
 O^\ast\!\left(
 \underbrace{1.40^n}_{N_{\rm small}}
 +\underbrace{1.77467^n}_{|\mathcal F_R(\mathcal D)|/\sqrt p}
 +\underbrace{1.67610^n}_{\sqrt{N_XN_A/p}}
 +\underbrace{1.77461^n}_{\sqrt{N_XN_Y/p}}
 \right)=O^\ast(1.77467^n).
\]
This proves the second pair of bounds in Theorem~\ref{thm:tradeoff}.

\paragraph{AI disclosure.}
The author proposed using a set system to compute an optimal value
with small success probability and amplifying this probability by
quantum search.
The author used GPT-6.2 Astra to find and optimize the
numerical parameters for the running-time and space--time bounds.
AI assistance was also used in drafting and revising the exposition.

\paragraph{Acknowledgement.} The author was supported by JSPS KAKENHI Grant Number JP25K24383 and MEXT Supporting Pioneering Research through AI for 1,000 Discovery challenges Program (SPReAD) Japan Grant Number JPMXP1726277558.

\bibliographystyle{alpha}
\bibliography{refs}

\appendix

\section{Applications}\label{app:applications}
We prove the \(O^*(1.70961^n)\) running-time bound in
Corollary~\ref{cor:applications}. For Minimum Set Cover, Directed
Feedback Arc Set, and One-Sided Crossing Minimization, we give
the recurrence for computing the cost of each part and show how
to combine the four costs in the search of Section~\ref{sec:four-path-search}. For Shortest
Common Superstring, we obtain the same bound by a reduction to
directed TSP on at most \(n+1\) cities.
\paragraph{Minimum Set Cover.}
Let \(U\) be a finite set of \(n\) elements. The input is a family
\(\mathcal S\) of subsets \(S_1,\ldots,S_m\) of \(U\). Each input set
\(S_j\) has a nonnegative, binary-encoded integer cost \(w_j\).
The task is to find a minimum-cost subfamily of \(\mathcal S\) whose union is \(U\),
or report that no cover exists.

For a subset \(Z\subseteq U\), let \(\operatorname{cov}(Z)\) be the
minimum cost of input sets whose union contains \(Z\). We use the weighted version of the recurrence used in~\cite{abboud-dp}
\begin{equation}\label{eq:app-cover-dp}
 \operatorname{cov}(\varnothing)=0,\qquad
 \operatorname{cov}(Z)=\min_{j:\,S_j\cap Z\ne\varnothing}
    \bigl(w_j+\operatorname{cov}(Z\setminus S_j)\bigr)
 \quad(Z\ne\varnothing).
\end{equation}

\smallskip\noindent\emph{The four-part search.}
We use the partition sizes and colorings of Section~\ref{sec:four-path-search}. For each
subfamily \(H\subseteq\mathcal S\) with \(|H|\le3\), the search
evaluates a partition \((A,B,C,D)\) of \(U\) by
\begin{equation}\label{eq:app-cover-four}
 \sum_{j:S_j\in H} w_j
 +\bigl[T_R^H(A)+T_{\rm small}^H(B)\bigr]
 +\bigl[T_{\rm small}^H(C)+T_{\rm small}^H(D)\bigr].
\end{equation}
Both tables use recurrence~\eqref{eq:app-cover-dp} with set costs
\[
 w_j^H=
 \begin{cases}
  0   & \text{if }S_j\in H,\\
  w_j & \text{if }S_j\notin H.
 \end{cases}
\]
The entry \(T_{\rm small}^H(Z)\) is the minimum cost of covering
\(Z\) at costs \(w_j^H\). We build this table once for each \(H\)
on all subsets of size at most \(\max\{b,c,d\}\).
For the table \(T_R^H\), let \(R\subseteq U\) be the set of red
elements and let \(\mathcal D\) be the set of allowed pairs of red
and blue counts from Section~\ref{sec:four-path-search}. The entry \(T_R^H(Z)\) is computed
by the same recurrence on the family
\[
 \mathcal F_R(\mathcal D)
 =\{Z\subseteq U:(|Z\cap R|,|Z\setminus R|)\in\mathcal D\}.
\]
In both tables the empty set has value zero, and a missing nonempty
entry or a minimum over no choices has value \(+\infty\).
Thus a set in \(H\) can be used in several parts at zero cost in
the tables. The first term of~\eqref{eq:app-cover-four} adds the
original cost of \(H\) once.

\smallskip\noindent\emph{Correctness.}
As in Section~\ref{sec:four-path-search}, \(R\subseteq U\) is the set of \(r\) red elements,
and \(r_A\) and \(r_X\) are the prescribed numbers of red elements
in \(A\) and \(X=A\cup B\), respectively.
For fixed \(R\), we minimize~\eqref{eq:app-cover-four} over all
\(H\subseteq\mathcal S\) with \(|H|\le3\) and all ordered partitions
\((A,B,C,D)\) of \(U\) of sizes \(a,b,c,d\) satisfying
\(|A\cap R|=r_A\) and \(|(A\cup B)\cap R|=r_X\).
We prove two statements:
\begin{enumerate}
\item\label{itm:cover-lower} For every \(R\) of size \(r\), the minimum is at least
      \(\operatorname{cov}(U)\).
\item\label{itm:cover-optimum} For \(R\) chosen uniformly among the \(r\)-element subsets of \(U\),
      the minimum equals \(\operatorname{cov}(U)\) with probability
      at least \(p\), where \(p\) is the lower bound from Section~\ref{sec:running-time}.
\end{enumerate}

\smallskip\noindent\emph{Proof of (\ref{itm:cover-lower}).}
For any finite value of~\eqref{eq:app-cover-four}, let \(\mathcal C\)
be the union of \(H\) and the four covers achieving the table entries
for \(A,B,C,D\). Then \(\mathcal C\) covers \(U\), and the value of~\eqref{eq:app-cover-four} is at least the cost of $\mathcal C$ which is at least \(\operatorname{cov}(U)\).

\smallskip\noindent\emph{Proof of (\ref{itm:cover-optimum}).}
Fix an optimal cover \(\mathcal C^\star\). Assign each \(x\in U\)
to the set \(S_j\in\mathcal C^\star\) of largest index containing
\(x\), and let \(G_j\subseteq S_j\) consist of the elements assigned
to \(S_j\). To form \(\pi^\star=(x_1,\ldots,x_n)\), append the
elements of \(G_j\) for each \(S_j\in\mathcal C^\star\) in increasing
order of \(j\), using any order within each \(G_j\).
Let \(A^\star,B^\star,C^\star,D^\star\) be the element sets of the
four consecutive parts of \(\pi^\star\) of sizes \(a,b,c,d\).

We show that the minimum of~\eqref{eq:app-cover-four} equals
\(\operatorname{cov}(U)\) with probability at least \(p\) over
the choice of \(R\). By (\ref{itm:cover-lower}), it remains to prove the upper bound.
For this bound, we use the partition
\((A^\star,B^\star,C^\star,D^\star)\) and construct a subfamily
\(H^\star\subseteq\mathcal C^\star\) with \(|H^\star|\le3\).

\smallskip\noindent\emph{Choice of \(H\).}
For each part \(P\in\{A^\star,B^\star,C^\star,D^\star\}\), we form
a cover of \(P\) from the input sets \(S_j\in\mathcal C^\star\)
for which \(G_j\) intersects \(P\).
Let \(H^\star\) consist of the input sets that belong to at least
two of the four covers. Since the elements of each \(G_j\) occur consecutively in \(\pi^\star\),
every \(S_j\in H^\star\) satisfies
\(G_j\cap\{x_a,x_{a+b},x_{a+b+c}\}\ne\varnothing\).
The sets \(G_j\) are pairwise disjoint, so \(|H^\star|\le3\).

\smallskip\noindent\emph{Using the four covers to bound the minimum.}
With \(H=H^\star\), we use the four covers of
\(A^\star,B^\star,C^\star,D^\star\) to prove that the minimum
of~\eqref{eq:app-cover-four} is at most \(\operatorname{cov}(U)\).
The partition must have the prescribed red counts, and the two
tables must contain the subsets needed to construct the four covers.
By~\eqref{eq:fixed-word-success}, a uniformly chosen \(r\)-element
subset \(R\) of \(U\) satisfies the following conditions simultaneously
with probability at least \(p\).
\[
 |A^\star\cap R|=r_A,\qquad
 |(A^\star\cup B^\star)\cap R|=r_X,\qquad
 \mathcal F_R(\mathcal D)\text{ supports }(x_1,\ldots,x_a).
\]

The first two conditions give the prescribed red counts.
For each \(P\in\{B^\star,C^\star,D^\star\}\), the table
\(T_{\rm small}^{H^\star}\) contains every subset of \(P\).
It remains to check that \(T_R^{H^\star}\) contains the subsets
needed for the cover of \(A^\star\).

In recurrence~\eqref{eq:app-cover-dp}, choose the sets in the cover
of \(A^\star\) in decreasing order of index. Each \(x\in A^\star\)
is first covered by its assigned set, because \(x\) was assigned
to the largest-index set in \(\mathcal C^\star\) containing \(x\).
After choosing \(S_j\), the uncovered elements of \(A^\star\) are
therefore exactly those assigned to sets of index less than \(j\).
Since \(\pi^\star\) lists the groups \(G_i\) in increasing order
of \(i\), the uncovered elements form a prefix set of
\((x_1,\ldots,x_a)\). By the third coloring condition,
\(T_R^{H^\star}\) contains every nonempty prefix set needed
to construct the cover of \(A^\star\).

Each of the four table entries is therefore at most the cost of
the corresponding cover under the costs used in the tables.
Each \(S_j\in H^\star\) has cost zero in the tables and contributes
\(w_j\) once to the first term of~\eqref{eq:app-cover-four}.
Each \(S_j\in\mathcal C^\star\setminus H^\star\) is used in at
most one of the four covers. Hence the minimum
of~\eqref{eq:app-cover-four} is at most the cost of
\(\mathcal C^\star\), namely \(\operatorname{cov}(U)\).
Together with (\ref{itm:cover-lower}), this proves that the minimum equals
\(\operatorname{cov}(U)\) with probability at least \(p\).

\paragraph{Directed Feedback Arc Set.}
Given a loopless directed graph on a vertex set \(V\) of size \(n\)
with nonnegative, binary-encoded integer arc weights, the task is to find a
minimum-weight set of arcs whose deletion makes the graph acyclic.
An arc is \emph{backward} in a vertex order if it goes from a later
vertex to an earlier vertex. Deleting all backward arcs leaves an
acyclic graph. Conversely, after any deletion that makes the graph
acyclic, the vertices admit an order in which every remaining arc
points forward. Thus the minimum deletion cost equals the minimum
total weight of backward arcs over all orders of \(V\).

\smallskip\noindent\emph{The minimum cost within one part.}
For distinct vertices \(u,v\), let \(c(u,v)\) be the weight of
the arc \(v\to u\) if present, and zero otherwise. For a family \(\mathcal F\) of vertex
subsets, let \(f_{\mathcal F}(Z)\) be the minimum weight of
backward arcs with both endpoints in \(Z\), over all orders of \(Z\)
whose nonempty prefix sets belong to \(\mathcal F\).
If \(v\) is last in \(Z\), the backward arcs incident to \(v\) are
exactly the arcs from \(v\) to \(Z\setminus\{v\}\). Minimizing over
the last vertex thus gives
\begin{equation}\label{eq:app-order-dp}
 f_{\mathcal F}(Z)=\min_{v\in Z}
 \left(f_{\mathcal F}(Z\setminus\{v\})
       +\sum_{u\in Z\setminus\{v\}}c(u,v)\right)
 \qquad(\varnothing\ne Z\in\mathcal F).
\end{equation}
The empty set has value zero, and a nonempty set outside
\(\mathcal F\) has value \(+\infty\).

As in Section~\ref{sec:four-path-search}, we divide an order of \(V\) into four consecutive
parts. Let \(A,B,C,D\) be the vertex sets of the four parts, with
prescribed sizes \(a,b,c,d\), respectively, where \(a+b+c+d=n\).

For the parts \(B,C,D\), we precompute a table \(T_{\rm small}\).
For every \(Z\subseteq V\) with \(|Z|\le\max\{b,c,d\}\), the entry
\(T_{\rm small}(Z)\) stores the minimum weight of backward arcs
over all orders of \(Z\). We compute these entries
by~\eqref{eq:app-order-dp} with
\(\mathcal F=\{Z\subseteq V:|Z|\le\max\{b,c,d\}\}\).

For the part \(A\), we build a table \(T_R\) for each coloring.
As in Section~\ref{sec:four-path-search}, \(R\subseteq V\) is the set of \(r\) red vertices,
and \(\mathcal D\) is the set of allowed pairs of red and blue counts.
The vertex sets allowed by the coloring form the family
\[
 \mathcal F_R(\mathcal D)
 =\{Z\subseteq V:(|Z\cap R|,|Z\setminus R|)\in\mathcal D\}.
\]
The entry \(T_R(Z)=f_{\mathcal F_R(\mathcal D)}(Z)\) stores the
minimum weight of backward arcs over orders of \(Z\) whose nonempty
prefix sets all belong to \(\mathcal F_R(\mathcal D)\).

\smallskip\noindent\emph{Combining the four parts.}
For a fixed coloring \(R\) and an ordered partition \((A,B,C,D)\),
the minimum total weight of backward arcs with both endpoints in
the same part is
\[
  T_R(A)+T_{\rm small}(B)+T_{\rm small}(C)+T_{\rm small}(D),
\]
where the order of \(A\) must be supported by \(\mathcal F_R(\mathcal D)\).
To compute the cost of backward arcs between parts, we use the
weighted form of the recurrence of Ambainis et al.~\cite{abi}.
If every vertex of \(L\) precedes every vertex of a disjoint set
\(Q\), the backward arcs between \(L\) and \(Q\) are exactly the arcs
from \(Q\) to \(L\). Their total weight is
\[
  \Delta(L,Q)=\sum_{u\in L}\sum_{v\in Q}c(u,v).
\]

As in Section~\ref{sec:four-path-search}, fix \(X=A\cup B\) with \(|X|=a+b\) and
\(|X\cap R|=r_X\), and put \(Y=V\setminus X\).
Since \(X\) precedes \(Y\), the cost \(\Delta(X,Y)\) is fixed.
We can therefore minimize independently over the split of \(X\)
into \(A,B\) and the split of \(Y\) into \(C,D\). Writing
\(B=X\setminus A\) and \(D=Y\setminus C\), the minimum for this
\(X\) is
\begin{equation}\label{eq:app-order-four}
\begin{aligned}
 \Delta(X,Y)
 &+\min_{\substack{A\subseteq X,\ |A|=a\\ |A\cap R|=r_A}}
    \bigl[T_R(A)+T_{\rm small}(B)+\Delta(A,B)\bigr]\\
 &+\min_{\substack{C\subseteq Y,\ |C|=c}}
    \bigl[T_{\rm small}(C)+T_{\rm small}(D)+\Delta(C,D)\bigr].
\end{aligned}
\end{equation}

In the actual search, we minimize~\eqref{eq:app-order-four} over all
\((a+b)\)-element sets \(X\subseteq V\) with \(|X\cap R|=r_X\). Every finite value of~\eqref{eq:app-order-four} is the backward-arc
weight of the vertex order obtained by concatenating the four
minimizing orders on \(A,B,C,D\).
Thus the minimum over \(X\) equals the optimum whenever the
prescribed red counts admit the partition of a fixed optimal order
and \(\mathcal F_R(\mathcal D)\) supports the order on its first part.

\paragraph{One-Sided Crossing Minimization.}
The input is a bipartite graph \(G=(L\cup V,E)\) with \(|V|=n\)
and a fixed left-to-right order \(\prec\) of the vertices in \(L\).
The vertices of \(L\) lie on an upper horizontal line in order
\(\prec\), and the vertices of \(V\) lie on a lower horizontal line.
Each edge is drawn as the line segment joining its endpoints.
The task is to choose a left-to-right order of \(V\) that minimizes
the number of pairs of edges that cross. When \(v\) comes before \(v'\), the number of
crossings between edges incident to \(v\) and to \(v'\) is
\[
 c(v,v')=\bigl|\{(u,u')\in L^2: uv\in E,\ u'v'\in E,\ u'\prec u\}\bigr|.
\]
Caroppo, Da Lozzo, and Di Battista~\cite{cdd} give the recurrence
for splitting the order of \(V\) at any fixed position. Since the crossing count is the sum of the pair costs \(c(v_i,v_j)\)
over \(i<j\), the recurrence~\eqref{eq:app-order-dp} and the
four-part search~\eqref{eq:app-order-four} apply just as in
Directed Feedback Arc Set.

\paragraph{Shortest Common Superstring.}
Given \(n\) finite strings, the task is to find a shortest string containing every
input string as a consecutive substring. We may assume without loss of
generality that the input strings are distinct and that no input
string is a substring of another. If at most one string remains,
the answer is immediate. For distinct input strings \(x,y\),
let \(o(x,y)\) be the length of the longest suffix of \(x\) that is
a prefix of \(y\). For an order \(x_1,\ldots,x_n\) of the input strings, start with
\(x_1\) and append each \(x_i\) for \(i=2,\ldots,n\), omitting
the first \(o(x_{i-1},x_i)\) characters because they already
appear at the end of \(x_{i-1}\). The resulting common superstring
has length
\[
  |x_1|+\sum_{i=2}^n \bigl(|x_i|-o(x_{i-1},x_i)\bigr).
\]
Khadiev et al.~\cite{khadiev-superstring} showed that
Shortest Common Superstring reduces to finding a maximum-weight
Hamiltonian path in the directed graph whose vertices are the
input strings and whose arc \(x\to y\) has weight \(o(x,y)\).
To reduce this path problem to minimum-weight directed TSP, we
add one city \(q\) and set
\[
 w(q,x)=|x|,\qquad w(x,y)=|y|-o(x,y),\qquad w(x,q)=0.
\]
The edge from \(q\) pays for the first string. Each edge \(x\to y\)
pays for the letters of \(y\) appended after the overlap. The tour
cost therefore equals the length of the merged string, so the problem is reduced to TSP on \(n+1\) cities.

\end{document}